\documentclass[preprint,12pt]{elsarticle}

\usepackage{amssymb}

\usepackage{color}

\usepackage{tabu}
\usepackage{xcolor}

\journal{Experimental Thermal and Fluid Science}

\begin{document}

\begin{frontmatter}



\title{Velocity fields of a bed-load layer under a turbulent liquid flow \tnoteref{label_note_copyright} \tnoteref{label_note_doi}}

\tnotetext[label_note_copyright]{\copyright 2016. This manuscript version is made available under the CC-BY-NC-ND 4.0 license http://creativecommons.org/licenses/by-nc-nd/4.0/}

\tnotetext[label_note_doi]{Accepted Manuscript for Experimental Thermal and Fluid Science, v. 78, p. 220-228, 2016, doi:10.1016/j.expthermflusci.2016.06.013}


\author{Marcos Roberto Mendes Penteado}
\ead{mmendes@fem.unicamp.br}
\author{Erick de Moraes Franklin\corref{cor1}}
\ead{franklin@fem.unicamp.br}

\cortext[cor1]{Corresponding author}

\address{Faculty of Mechanical Engineering - University of Campinas - UNICAMP\\
Rua Mendeleyev, 200 - Campinas - SP - CEP: 13083-970\\
Brazil}

\begin{abstract}
The transport of sediments by a fluid flow is commonly found in nature and in industry. In nature, it is found in rivers, oceans, deserts, and other environments. In industry, it is found in petroleum pipelines conveying grains, in sewer systems, and in dredging lines, for example. This study investigates experimentally the transport of the grains of a granular bed sheared by a turbulent liquid flow. In our experiments, fully developed turbulent water flows were imposed over a flat granular bed of known granulometry. Under the tested conditions, the grains were transported as bed load, i.e., they became entrained by rolling and sliding over each other, forming a moving granular layer. The present experiments were performed close to incipient bed load, a case for which experimental data on grains velocities are scarce. Distinct from previous experiments, an entrance length assured that the water stream over the loose bed was fully developed. At different water flow rates, the moving layer was filmed using a high-speed camera, and the grains' displacements and velocities were determined by post-processing the images with a numerical code developed in the course of this study. The bed-load transport rate was estimated and correlated to the water flow conditions.
\end{abstract}


\begin{keyword}
Sediment transport \sep bed load \sep turbulent flow \sep individual grains

\end{keyword}

\end{frontmatter}



\section{Introduction}

The transport of sediments by a fluid flow is directly related to the shear forces exerted by the fluid on the granular bed. When the shear forces are very strong in relation to the grains' weight, the granular bed is fluidized and grains are entrained as a suspension. Conversely, when the ratio between the shear forces and the grains' weight is moderate, the grains are entrained by the fluid flow as a mobile granular layer, known as bed load, which remains in contact with the fixed part of the granular bed. Within the bed-load layer, grains become entrained by rolling and sliding over each other, or by effectuating small jumps.

The transport of sediments as bed load by a turbulent liquid flow is commonly found in nature and in industry. For example, it can be found in rivers, oceans, petroleum pipelines conveying grains, sewer systems, and dredging lines. In practical situations, the bed load accounts for a considerable proportion of the transport rate, and therefore, it is of interest for many environmental and industrial applications. Although of importance, the problem is not fully understood and remains open. Some of the major difficulties related to the problem are the feedback mechanisms. For example, the water stream is responsible for the solid discharge, which, in its turn, modifies the morphology of the bed through erosion and sedimentation processes \cite{Franklin_4, Franklin_6}. Another example is the modification of the fluid flow by momentum transfers from the fluid to the moving grains, and from the latter to the fixed part of the bed, known as the feedback effect \cite{Franklin_8}.

To determine the bed-load transport rate, two dimensionless groups are necessary, usually taken as the Shields number, $\theta$, and the particle Reynolds number, $Re_*$. The Shields number is the ratio between the shear force caused by the fluid and the grains' weight, and the particle Reynolds number is the Reynolds number at the grain scale \cite{Raudkivi_1}. They are given by Eqs. \ref{equation1} and \ref{equation2}, respectively:
	
\begin{equation}
\theta = \frac{\tau}{(\rho_{s}-\rho)gd}
\label{equation1}
\end{equation}

\begin{equation}
Re_* = \frac{u_*d}{\nu}
\label{equation2}
\end{equation}

\noindent Here $\tau$ is the shear stress caused by the fluid on the granular bed, $d$ is the grain diameter, $g$ is the gravitational acceleration, $\nu$ is the kinematic viscosity, $\rho$ is the specific mass of the fluid, and $\rho_{s}$ is the specific mass of the grain. In the case of two-dimensional turbulent boundary layers, the shear stress is $\tau\,=\,\rho u_*^2$ , where $u_*$ is the shear velocity. The mean velocity $u$ in the overlap region is given by

\begin{equation}
u^+\,=\,\frac{1}{\kappa} ln \left( \frac{z}{z_0} \right) \,=\,\frac{1}{\kappa} ln(z^+) +B,
\label{mean_velocity}
\end{equation}

\noindent where $\kappa\,=\,0.41$ is the von K\'arm\'an constant, $z_0$ is the roughness length, $u^+\,=\,u/u_*$ is a dimensionless velocity, $z^+\,=\,zu_*/\nu$ is the vertical distance normalized by the viscous length, and $B$ is a constant. In Eq. \ref{mean_velocity}, the second $\left( \sim ln \left( z/z_0 \right) \right)$ and third $\left( \sim ln(z^+) \right)$ terms are equivalent, the second being generally employed for hydraulic rough regimes and the third for hydraulic smooth regimes, for which $B=5.5$.

Some previous studies were devoted to the kinematic properties of moving grains within a bed-load layer under liquid flows. Fernandez Luque and van Beek (1976) \cite{Fernandez_luque} presented an experimental study on the average motion of individual grains and bed-load transport rates. The experiments were performed in a $8\, m$ long, $0.20\, m$ high and $0.10\, m$ wide inclinable closed-conduit channel, where they imposed different water flow rates over different granular beds. The authors measured the bed-load transport rate, deposition rate, mean grain velocity, and displacement length of grains. They proposed that the mean longitudinal velocity of grains $v_x$ is given by $v_x = 11.5 (u_* - 0.7u_{*,th})$, where $u_{*,th}$ is the shear velocity corresponding to the threshold shear stress.

Charru et al. (2004) \cite{Charru_1} presented an experimental study on the dynamics of a granular bed sheared by a viscous Couette flow in the laminar regime. The experimental results concerned the displacement of individual grains (velocities, durations, and lengths) and the surface density of the moving grains. For a given shear stress caused by the fluid, at a given flow rate, and an initially loose bed, Charru et al. (2004) \cite{Charru_1} showed that the surface density of moving grains decays while their velocity remains unchanged. They proposed that this decay is due to an increase in bed compactness, caused by the rearrangement of grains, known as armoring, which leads to an increase in the threshold shear rate for the bed load. They found that the velocity of individual grains is approximately given by the shear rate times the grain diameter times a constant factor equal to $0.1$, and that the duration of displacements is approximately given by $15$ times the settling time, considered to be the grain diameter divided by the settling velocity of a single grain. Charru et al. (2009) \cite{Charru_4} presented experiments on laminar flow in which PIV (Particle Image Velocimetry) was used to measure the velocity profiles of the fluid inside the mobile granular layer, as well as the grains' displacements, by using a background subtraction technique. They proposed correlations between the displacements of grains and the fluid flow, applicable in the laminar regime.

Lajeunesse et al. (2010) \cite{Lajeunesse_1} presented an experimental study on the motion of individual grains of a bed-load layer over a flat granular bed. Free-surface turbulent water flows in steady state regime, with Reynolds numbers based on the water depth between $1500$ and $6000$, were imposed over different granular beds. The granular beds consisted of three populations of grains, employed separately: quartz grains with median diameters of $1.15\,mm$, $2.24\,mm$, and $5.5\,mm$, corresponding to $12\,\leq\, Re_* \,\leq\,500$. The displacements of individual grains were filmed using a high-speed camera, and the grains velocities, displacement lengths, and durations of flights were determined from the acquired images. The authors found that the distributions of longitudinal and transverse grain velocities follow, respectively, a decreasing exponential law and a Gaussian law, that the surface density of moving grains varies with $\theta - \theta_{th}$, and that the grains velocity and flight length vary with $\theta^{1/2}-\theta_{th}^{1/2}$, where $\theta_{th}$ is the Shields number corresponding to the threshold shear stress. They also found that the flight duration scales with the settling velocity of a single grain.

In the last few decades, many works were devoted to correlating the bed-load transport rate as a function of the fluid flow. Meyer-Peter and  M\"{u}ller (1948) \cite{Mueller}, applying their data from exhaustive experimental work, proposed that $\phi_B\,=\,a(\theta - \theta_{th})^{3/2}$, where $a=8$ if both form drag (due to ripples) and skin friction are considered, and $a=4$ if only skin friction is considered \cite{Wong_parker}. $\phi_B\,=\,q_B((S-1)gd^3)^{-1/2}$ is the normalized volumetric bed-load transport rate, where $S=\rho_s / \rho$ is the ratio between the specific masses and $q_B$ is the volumetric bed-load transport rate by unit of width. Bagnold (1956) \cite{Bagnold_3} proposed that $\phi_B\,=\,\eta \theta^{1/2}(\theta-\theta_{th})$, where $\eta$ is given by $\eta\,\approx\,A\sqrt{2\mu_s/(3C_D)}$, $A$ is a constant that depends on the Reynolds number \cite{Bagnold_3}, $\mu_s$ is the friction between grains, and $C_D$ is the drag factor for the grains. Based on energetic considerations, Bagnold (1973) \cite{Bagnold_4} proposed that $i_b \tan \alpha = e_b \omega$, where $i_b$ is the immersed bed-load transport rate, $\omega$ is the flow power per unit area, $\tan \alpha$ is the dynamic friction coefficient, and $e_b$ is the efficiency of bed-load transport (proportion of the flow power dissipated within the bed-load transport). Lettau and Lettau (1978) \cite{Lettau} proposed that $\phi_B\,=\,\xi\theta (\sqrt{\theta} - \sqrt{\theta_{th}})$, where $\xi\,=\,C_L((S-1)gd)^{-3/2}\rho /g$ and $C_L$ is a constant to be adjusted.

Abrahams and Gao (2006) \cite{Abrahams_gao} developed a bed-load transport equation valid for open-channel turbulent flows in rough regime. The obtained equation, derived from Bagnold's energy approach \cite{Bagnold_3} and based on exhaustive experimental data, is $i_b=\omega G^{3.4}$, where $G = 1-(\theta / \theta_{th})$. This equation was proposed for both bed-load and sheet-flow regimes by replacing the dynamic friction coefficient of Bagnold (1973) \cite{Bagnold_4} by a stress coefficient that takes into account both the grains contact and the fluid drag.

According to his measurements of aquatic dunes under turbulent liquid flows, Franklin (2008) \cite{Franklin_3} proposed that $\phi_B\,=\,12Re_* \left(\theta - \theta_{th} \right)^{3}$. As opposed to previous expressions, the proposed relation has an explicit dependence on $Re_*$. The author argued that the bed-load transport rate must vary with the  type of turbulent boundary layer, i.e., hydraulically smooth or hydraulically rough. Using the same argument, Franklin and Charru (2011) \cite{Franklin_9} proposed that $\phi_B\,=\,34Re_s \left(\theta - \theta_{th} \right)^{2.5}$, where $Re_s\,=\,U_sd/\nu$ is the Reynolds number based on the settling velocity of a single grain, $U_s$.

Gao (2012) \cite{Gao} compared the equation proposed by Abrahams and Gao (2006) \cite{Abrahams_gao} with exhaustive experimental data and other bed-load equations. For this compilation, Gao (2012) \cite{Gao} used a data set from 264 flume or closed-conduit experiments not used in obtaining the Abrahams and Gao (2006) \cite{Abrahams_gao} equation. The author found that the Abrahams and Gao (2006) \cite{Abrahams_gao} equation has the best predictive capacity among the tested bed-load equations.

Recently, Franklin et al. (2014) \cite{Franklin_8} measured the velocity profiles of turbulent water flows over fixed and loose granular beds of the same granulometry using PIV. They quantified the perturbation caused solely by the effect of the bed load on the turbulent stream, known as the feedback effect. Their experimental device had an entrance length of $40$ hydraulic diameters upstream of the loose bed. In the entrance length, a static bed of same thickness and granulometry of the loose bed was fixed on the bottom of the channel, assuring that the water stream was fully developed over the loose granular bed. The time scale of the experiments were such that no ripples were observed in the course of tests.

The objective of the study is to, first, determine experimentally the displacement and velocity fields of the moving grains within the bed-load layer in a liquid, and, second, as a consequence, to be able to estimate the bed-load transport rate from these. In order to achieve this, a granular bed was filmed using a high-speed camera and the images were post-processed using scripts developed by the authors. The present experiments were performed close to incipient bed load, a case for which experimental data on grains velocities are scarce, and the granular bed remained flat in the course of all tests. Distinct from previous experiments, an entrance length of $40$ hydraulic diameters assured that the water stream over the loose bed was fully developed.

\section{Experimental Device}
       
\begin{figure}[!ht]
  \begin{center}
    \includegraphics[width=0.90\columnwidth]{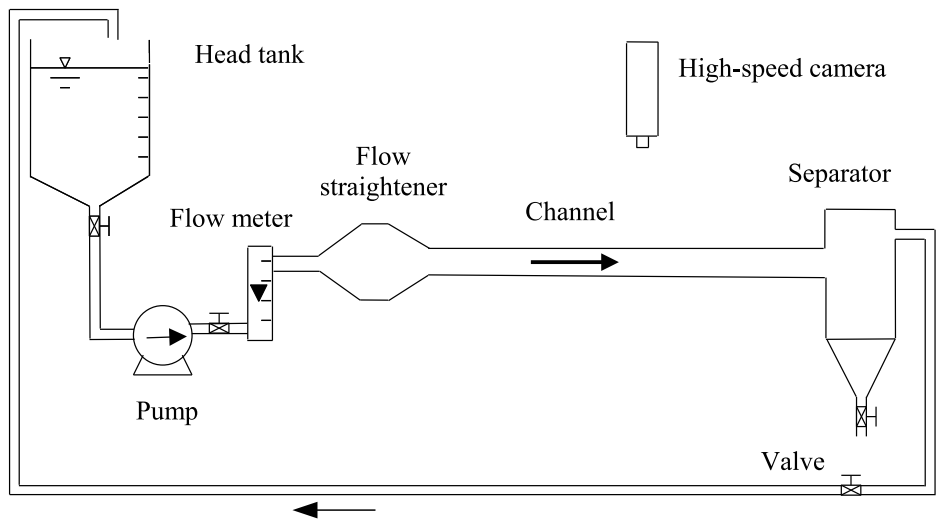}
    \caption{Scheme of the experimental loop.}
    \label{fig:loop}
  \end{center}
\end{figure}

The experimental device consisted of a water reservoir, a progressive pump, a flow straightener, a $5\,m$ long channel, a settling tank, and a return line. The flow straightener was a divergent-convergent nozzle filled with $d=3\,mm$ glass spheres, the function of which was to homogenize the flow profile at the channel inlet. The channel had a rectangular cross section ($160\,mm$ wide by $50\,mm$ high) and was made of transparent material. Figure \ref{fig:loop} shows the scheme of the experimental loop. The channel test section was $1\,m$ long and started $40$ hydraulic diameters ($3\,m$) downstream of the channel inlet. A fixed granular bed consisting of glass spheres glued on the surface of PVC plates was inserted in the channel section in which the flow is developed, assuring that the turbulent flow was completely developed in the test section. In the test section, the grains were deposed and formed a loose granular bed of the same thickness ($7\,mm$) as the fixed bed. Glass spheres with a specific mass of $\rho_s\,=\,2500\,kg/m^3$ and diameter ranging from $d\,=\,400\,\mu m$ to $d\,=\,600\,\mu m$ were employed to form both the fixed and loose granular beds. Prior to each test, the loose granular bed was smoothed and levelled and, in order to facilitate the post-treatment of images, $5\%$ of the granular bed consisted of black glass spheres, of the same granulometry and material.

The tests were performed at ambient conditions, i.e., atmospheric pressure of $1\,atm$ and temperature of approximately $25^{\circ}C$. The water flowed in a closed loop driven by the pump from the reservoir, through the channel and the grain separator, and back to the reservoir. The water flow rate was controlled by changing the excitation frequency of the pump and was measured with an electromagnetic flow meter, accurate to within $0.5 \%$ of the measured value. The test nominal flow rates were in the range $4.8\,m^3/h\,\leq\,Q\,\leq 8.1\,m^3/h$. The cross-section mean velocities were in the range $0.19\,m/s\,\leq\,\overline{U}\,\leq\,0.29\,m/s$ and the Reynolds number $Re\,=\, \overline{U}2H_{gap}/\nu$ in the range $1.6\cdot 10^4\,\leq\,Re\,\leq 2.7\cdot 10^4$, where $H_{gap}$ is the distance from the granular bed to the top wall. These upper and lower bounds were fixed to allow bed load close to incipient conditions and, at the same time, avoid the fast formation of ripples on the granular bed. Under these conditions, the moving layer occurred in a thickness corresponding to one grain diameter. Franklin et al. (2014) \cite{Franklin_8} measured water flow fields in the same experimental device over similar moving beds. The water flow profiles presented in Franklin et al. (2014) \cite{Franklin_8} were assumed to be valid for the present experiments and were used in this study.

\begin{figure}[!ht]
  \begin{center}
    \includegraphics[width=0.90\columnwidth]{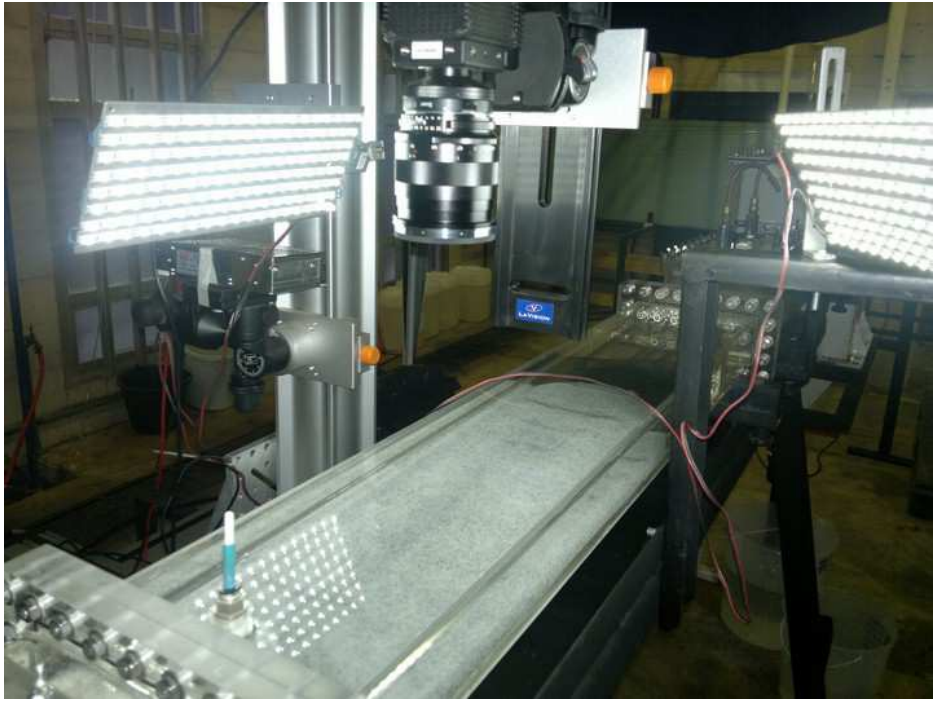}
    \caption{Test section.}
    \label{fig:fotobancada}
  \end{center}
\end{figure}

A high-speed CCD camera with a resolution of $1280\,px\,\times\,1024\,px$ at frequencies that can reach 1000 Hz was employed to obtain the displacements of grains. For the present tests, the frequency was adjusted between $90\,Hz$ and $500\,Hz$. A computer was used to control the frequencies and exposure times of the high-speed camera and to store the acquired images. In order to provide the necessary light for low exposure times while avoiding beating between the light source and the camera frequency, approximately $200$ LED (Light-Emitting Diode) lamps were attached to plates and branched to a continuous current source. For all tests, the number of acquired images was $1415$, the total field was $3250\,mm^2$, and a Makro-Planar lens of $50\,mm$ focal distance was used. Figure \ref{fig:fotobancada} shows the camera and the LED plates. The calibration process was performed once, after which the camera did not need adjustments. This allowed the conversion from pixels to the metric system ($mm$).

\section{Image Treatment}
  
A numerical code was developed in order to determine from the high-speed movies the displacements and velocities of moving grains in an Eulerian framework. Basically, the instantaneous displacement fields of the black spheres were computed by comparing pairs of frames and finding the displacements of the moving black spheres. The instantaneous velocity fields of black spheres were then computed by multiplying the displacement fields by the acquiring frequency. As black spheres constituted $5\%$ of the bed, the total number of moving grains was estimated from the number of moving black spheres, and the displacement and velocity fields of the black spheres were considered a representative sample of those of all the grains. The general aspects of the developed code are described next.

The image treatment comprises two main steps. The first is to identify the regions where black spheres were moving. This is achieved by binarizing the frames, employing a suitable intensity threshold, and by subtracting two consecutive frames. The resulting file is a matrix where $0$ corresponds to black and $1$ to white, and therefore, the code needs to consider only the regions for which the matrix elements are $1$.

\begin{figure}[!ht]
  \begin{center}
    \includegraphics[width=0.95\columnwidth]{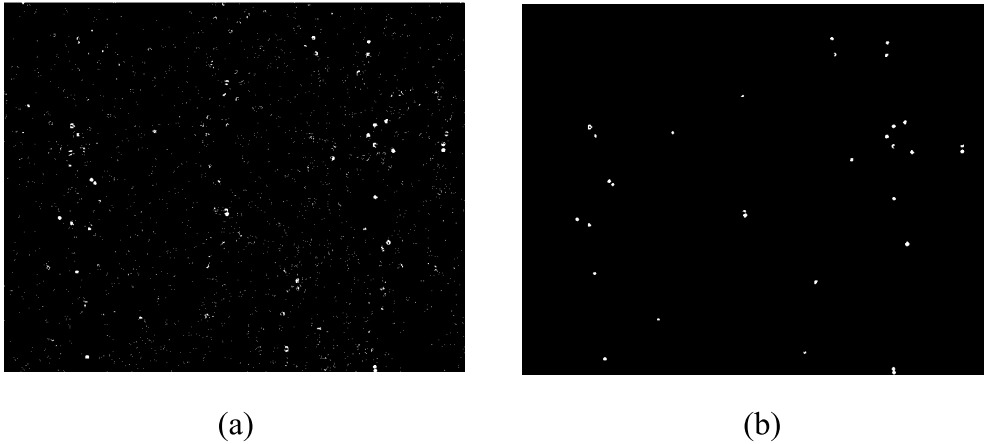}
    \caption{Subtraction of two consecutive frames. (a) Before filtering. (b) After filtering.}
    \label{fig:imagesubtraction}
  \end{center}
\end{figure}

Although a constant light source was used, small changes in illumination occurred between frames. This appears in the frames subtraction as noise: white regions with an area of the order of $1\, px^2$, which is one order of magnitude smaller than the grain diameters and the grain displacements, both of approximately $50\, px^2$. This is shown in Fig. \ref{fig:imagesubtraction}(a), where the larger white regions correspond to the displacements of the black spheres, while the smaller white regions correspond to noise due to variations in light intensity. To remove the noise, a median filter \cite{Sonka} and an area filter were applied. The area filter works based on the expected area of grain displacements: all the connected white regions with an area smaller than a threshold value ($5\, px^2$ in our experiments) are removed, i.e., changed from $1$ to $0$ in the binarized matrix. The result after filtering can be seen in Fig. \ref{fig:imagesubtraction}(b), where only the white regions corresponding to the displacements are present.

The second step is to compute the displacement distances of moving black spheres between two consecutive frames. This is achieved by identifying the moving black spheres in each frame of a pair, and then computing the location of their centroids. Finally, the subtraction of the corresponding centroids' positions (the code searches the nearest centroids within an assigned radius value) gives the instantaneous displacements. By multiplying the displacements by the acquiring frequency, we obtain the velocities.

\begin{figure}[!ht]
  \begin{center}
    \includegraphics[width=0.65\columnwidth]{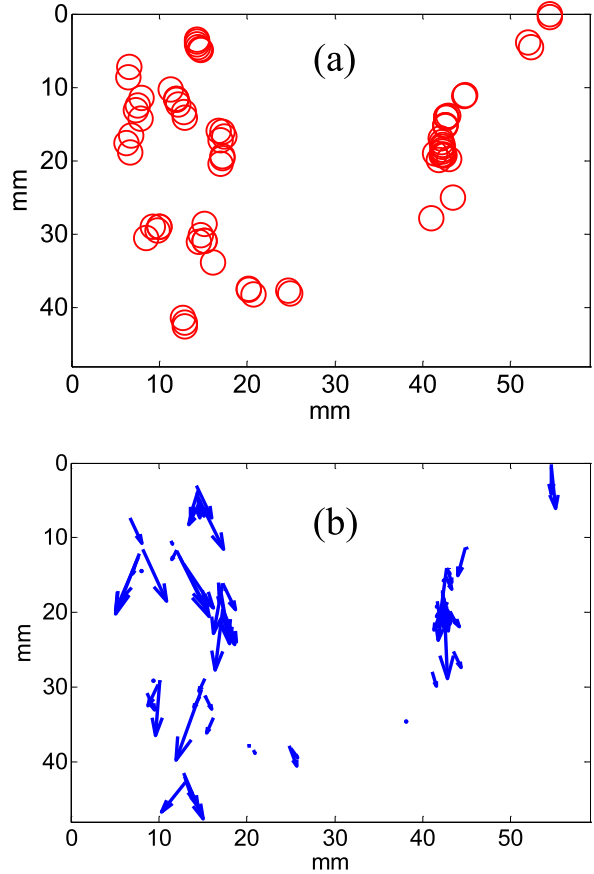}
    \caption{Subtraction of the computed centroids of consecutive frames. (a) Displacements of particles. (b) Velocities of particles.}
    \label{fig:displacements}
  \end{center}
\end{figure}

Figure \ref{fig:displacements}(a) shows the centroids of moving black spheres in a 10-frame sequence, allowing the individual displacements to be observed. Figure \ref{fig:displacements}(b) shows the velocity fields corresponding to the same frames (superposed).

The software Motion Studio was used to track individual grains in a Lagragian framework in some of the experiments.

\begin{figure}[!ht]
	\begin{center}
		\includegraphics[scale=0.8]{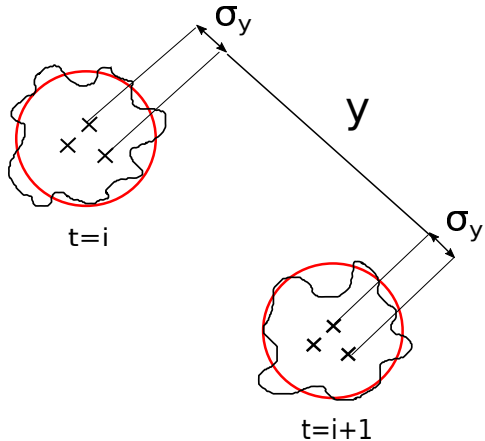}
		\caption{Uncertainty related to the determination of the centroids.}
		\label{fig:displac5a}
	\end{center}
\end{figure}

\begin{figure}[!ht]
	\begin{center}
		\includegraphics[scale=0.4]{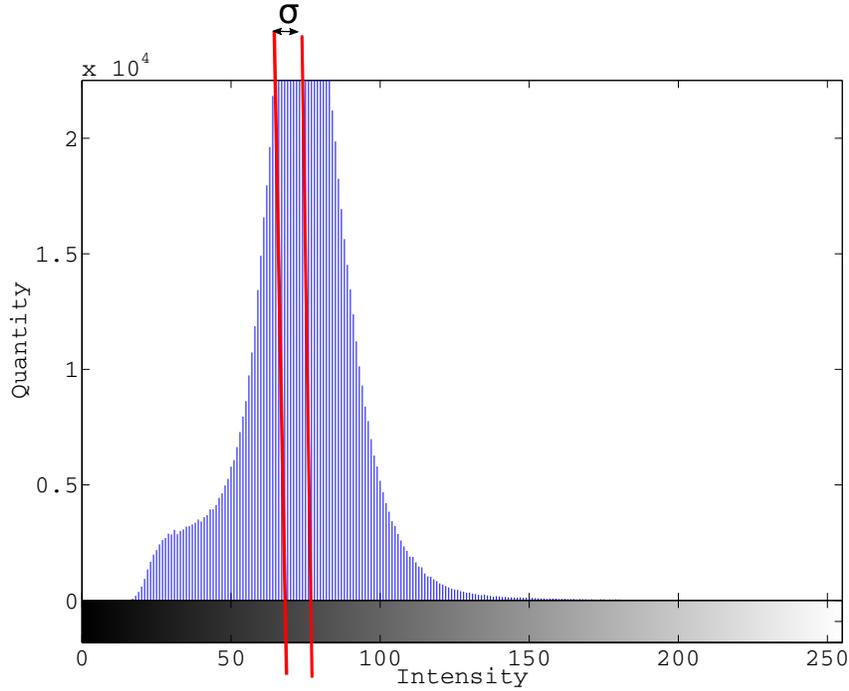}
		\caption{Histogram of light intensity}
		\label{fig:displac5d}
	\end{center}
\end{figure}

There are three main sources of uncertainties in the determination of the grains displacements. The first is given by the computation of the centroids by the numerical code. This is related to the images resolution, as can be observed in Fig. \ref{fig:displac5a}. The second is given by differences in images contrasts due to variations of the light intensity during the experiments. Figure \ref{fig:displac5d} shows, as an example, the histogram of light intensity for one frame. The last one is given by the $pixel-mm$ calibration. Combining all these three sources, the uncertainty of the computed displacements is of approximately 30$\mu m$.

\section{Results}

\subsection{Eulerian framework}

Franklin et al. (2014) \cite{Franklin_8} measured the water profiles over moving beds the grains of which consisted of glass spheres with $\rho_s=2500 kg/m^3$ and a diameter ranging from $d\,=\,300\,\mu m$ to $d\,=\,425\,\mu m$. The shear velocity $u_*$, particle Reynolds number $Re_*$, and Shields number $\theta$ presented in Tab. \ref{table:table1} were computed using Franklin et al.'s (2014) \cite{Franklin_8} data. Table \ref{table:table1} also presents the water flow rate $Q$, the mean displacement velocities in directions $x$ and $y$, $v_x$ and $v_y$, respectively, and the normalized mean displacement velocity in direction $x$, $v_x^{ad}$, for different test runs. In this table, $x$ corresponds to the longitudinal and $y$ to the transverse direction. The dimensionless mean displacement velocity $v_x^{ad}$ was obtained by normalizing $v_x$ by the shear rate and the typical grain diameter, $d$, so that $v_x^{ad} = \nu v_x/(u_*^2d)$.

Some of the test runs presented in Tab. \ref{table:table1} have the same water flow rate. Although the procedure for bed preparation was always the same, the discrete nature of the bed may result in initial beds with different compactnesses, so that armoring varies between runs. This may result in different threshold Shields numbers, and, for this reason, we performed more than one time some runs with the same water flow rates.

\begin{figure}[!ht]
   \begin{minipage}[c]{.49\linewidth}
    \begin{center}
     \includegraphics[width=0.9\columnwidth]{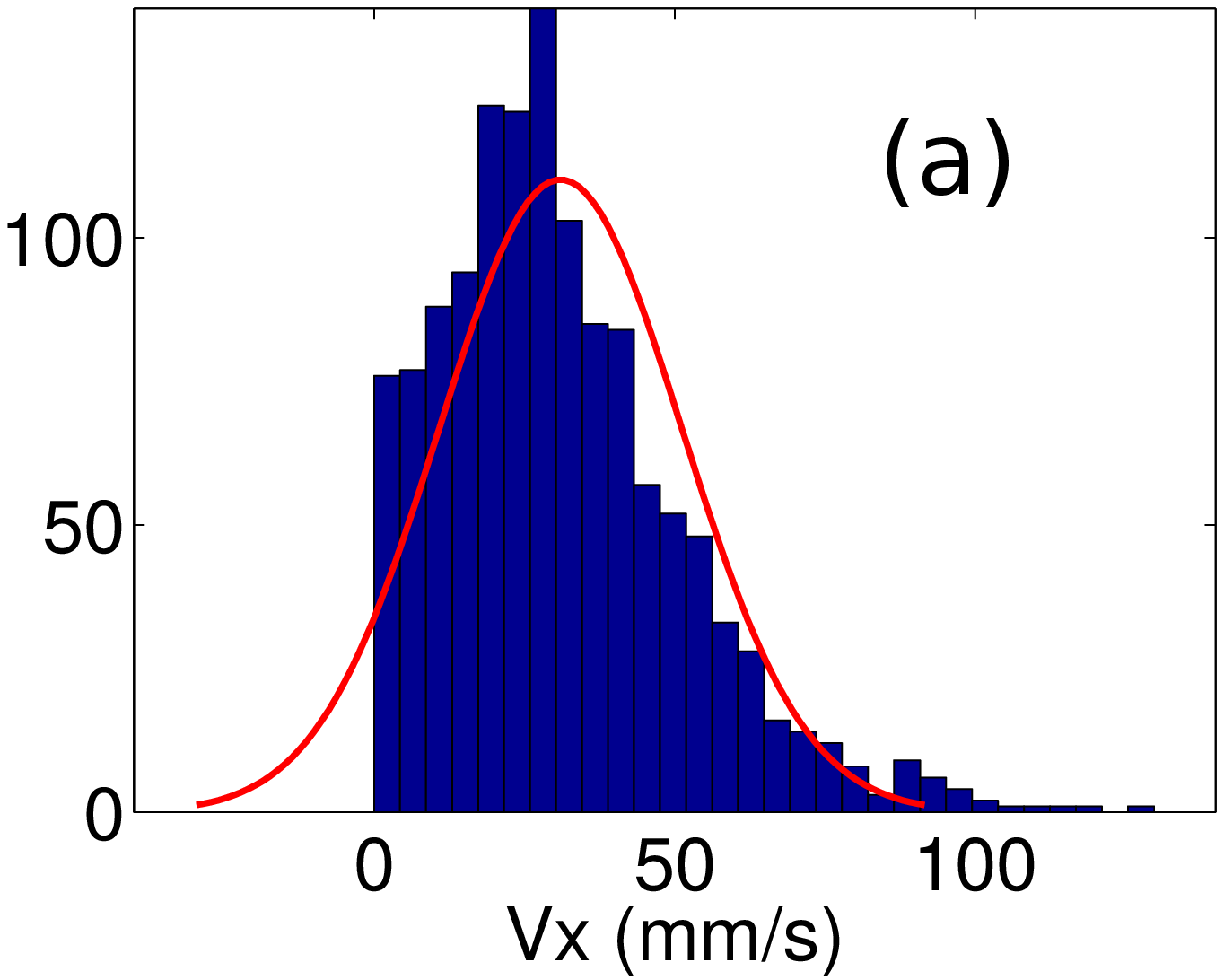}
    \end{center}
   \end{minipage} \hfill
   \begin{minipage}[c]{.49\linewidth}
    \begin{center}
      \includegraphics[width=0.9\columnwidth]{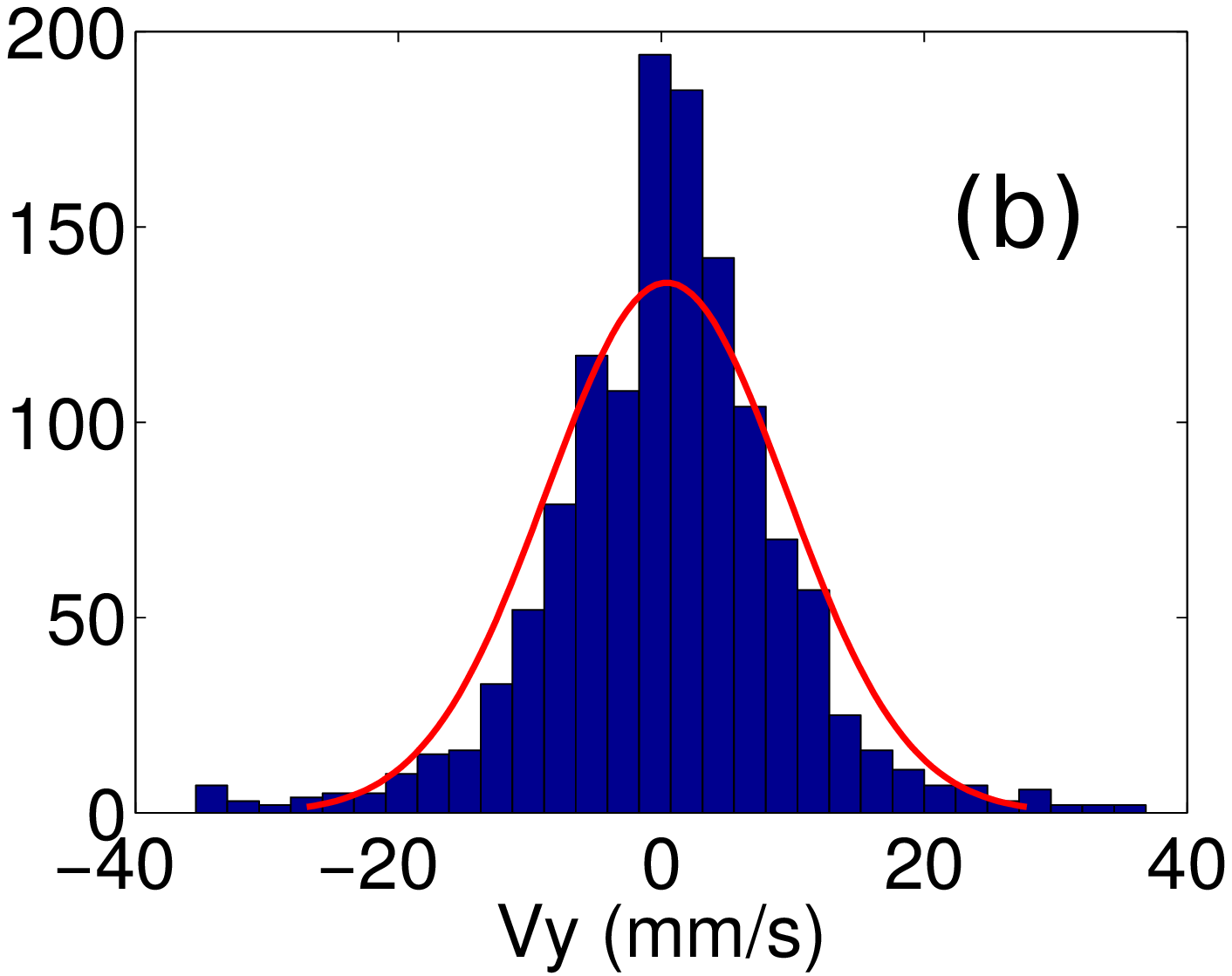}
    \end{center}
   \end{minipage}
\caption{Histograms for run 19, with a superimposed Gaussian function (a) $v_x$. (b) $v_y$.}
\label{histogram_2g}
\end{figure}

\begin{figure}[!ht]
   \begin{minipage}[c]{.49\linewidth}
    \begin{center}
     \includegraphics[width=0.9\columnwidth]{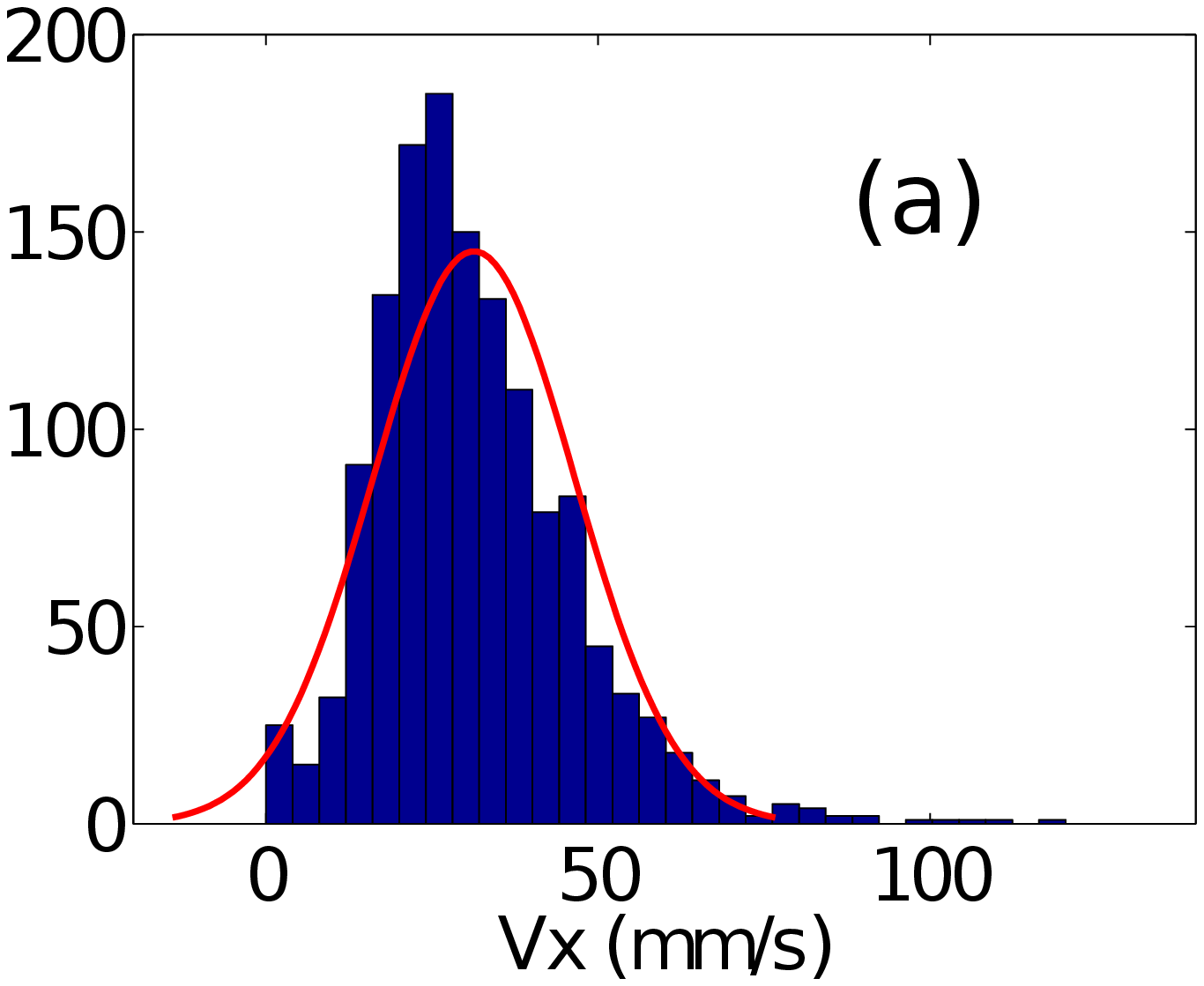}
    \end{center}
   \end{minipage} \hfill
   \begin{minipage}[c]{.49\linewidth}
    \begin{center}
      \includegraphics[width=0.9\columnwidth]{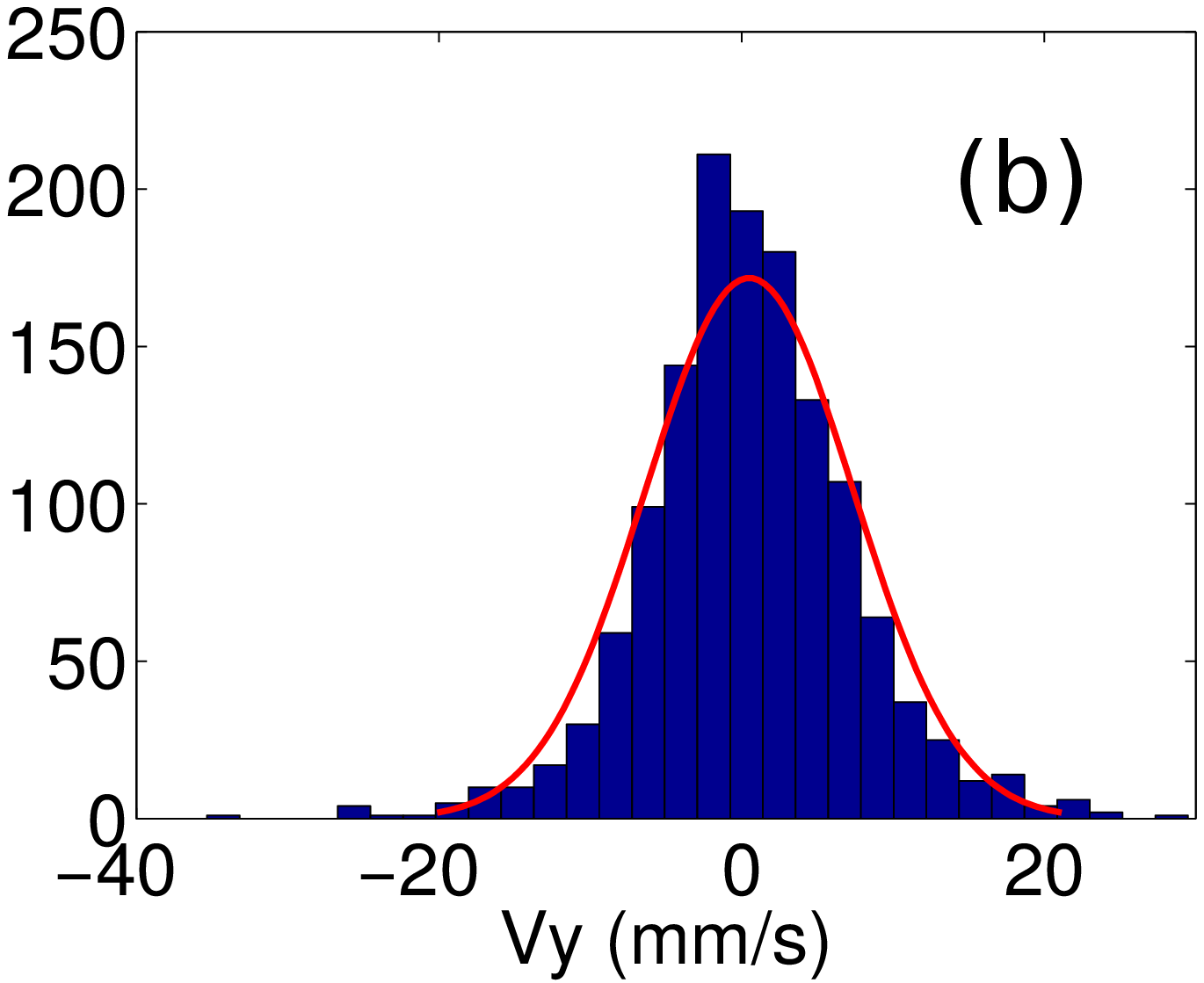}
    \end{center}
   \end{minipage}
\caption{Histograms for run 31, with a superimposed Gaussian function (a) $v_x$. (b) $v_y$.}
\label{histogram_4c}
\end{figure}

Figures \ref{histogram_2g} and \ref{histogram_4c} present the histograms of $v_x$ and $v_y$ for runs 19 and 31, respectively, with superimposed Gaussian functions. These histograms, which are representative of all the test runs, show that the larger frequencies of occurrence are concentrated around the mean values presented in Tab. \ref{table:table1}. The width of the Gaussian function characterizes the order of magnitude of the velocities fluctuations along the $x$ axis. According to Figs. \ref{histogram_2g}(a) and \ref{histogram_4c}(a), $v_x$ may be considered as the characteristic velocity of grains in longitudinal direction. Figures \ref{histogram_2g}(b) and \ref{histogram_4c}(b) show a Gaussian function centered at the origin, which is consistent with a particle trajectory mainly oriented along flow direction, i.e., that $v_y \approx 0$.

In the present experiments, we found that $v_x^{ad}\,=\, ord(0.1)$, where $ord$ denotes \textit{order of magnitude}, in agreement with Charru et al.'s (2004) results \cite{Charru_1}, although their experiments addressed viscous flows. This agreement may be explained by the fact that the present experiments were performed in a hydraulic smooth regime or close to the lower bound of the transitional regime ($Re_*\leq 11$).

\begin{table}[ht]
\caption{Test run, water flow rate $Q$, shear velocity $u_*$, Shields number $\theta$, particle Reynolds number $Re_*$, mean displacement velocity in $x$ direction $v_x$, mean displacement velocity in $y$ direction $v_y$, and normalized mean displacement velocity in $x$ direction $v_x^{ad}$.}
\label{table:table1}
\centering
\begin{tabu}{c c c c c c c c}  
\hline\hline
run & $Q$ & $u_*$ & $\theta$ & $Re_*$ & $v_x$ & $v_y$ & $v_x^{ad}$ \\
$\cdots$ & $m^3/h$ & $m/s$ & $\cdots$ & $\cdots$ & $mm/s$ & $mm/s$ & $\cdots$ \\ [0.5ex] 
\hline 
1 & 4.8 & 0.0102 & 0.014 & 5 & 12.8 & 0.3 & 0.25\\ 
2 & 5.7 & 0.0136 & 0.025 & 7 & 32.3 & 0.2 & 0.35\\ 
3 & 6.2 & 0.0154 & 0.032 & 8 & 51.4 & 1.4 & 0.43\\ 
4 & 6.2 & 0.0154 & 0.032 & 8 & 22.2 & 0.0 & 0.19\\ 
5 & 6.3 & 0.0157 & 0.033 & 8 & 48.0 & -1.2 & 0.39\\ 
6 & 6.3 & 0.0158 & 0.034 & 8 & 31.1 & 0.3 & 0.25\\ 
7 & 6.4 & 0.0160 & 0.035 & 8 & 27.4 & 0.2 & 0.21\\ 
8 & 6.4 & 0.0160 & 0.035 & 8 & 41.5 & 1.3 & 0.33\\ 
9 & 6.5 & 0.0165 & 0.037 & 8 & 40.1 & -1.4 & 0.29\\ 
10 & 6.5 & 0.0165 & 0.037 & 8 & 31.7 & 0.2 & 0.23\\ 
11 & 6.5 & 0.0165 & 0.037 & 8 & 28.2 & -0.7 & 0.21\\
12 & 6.6 & 0.0169 & 0.039 & 8 & 81.3 & -3.69 & 0.57\\ 
13 & 6.6 & 0.0169 & 0.039 & 8 & 49.5 & 0.51 & 0.34\\ 
14 & 6.6 & 0.0169 & 0.039 & 8 & 51.1 & 1.27 & 0.35\\ 
15 & 6.7 & 0.0173 & 0.041 & 9 & 73.0 & -7.5 & 0.49\\ 
16 & 6.7 & 0.0173 & 0.041 & 9 & 54.7 & 2.1 & 0.36\\ 
17 & 6.8 & 0.0177 & 0.042 & 9 & 33.3 & 0.2 & 0.21\\
18 & 6.9 & 0.0180 & 0.044 & 9 & 33.3 & 0.3 & 0.21\\
19 & 7.0 & 0.0183 & 0.046 & 9 & 31.0 & 0.4 & 0.18\\ 
20 & 7.0 & 0.0183 & 0.046 & 9 & 34.9 & 0.3 & 0.21\\ 
21 & 7.0 & 0.0184 & 0.046 & 9 & 50.5 & -2.9 & 0.30\\ 
22 & 7.1 & 0.0188 & 0.048 & 9 & 51.9 & -0.7 & 0.29\\ 
23 & 7.1 & 0.0188 & 0.048 & 9 & 76.8 & -0.8 & 0.44\\ 
24 & 7.2 & 0.0192 & 0.050 & 10 & 52.9 & 0.1 & 0.29\\ 
25 & 7.3 & 0.0194 & 0.051 & 10 & 81.8 & 1.76 & 0.44\\ 
26 & 7.3 & 0.0194 & 0.051 & 10 & 55.9 & 1.55 & 0.30\\ 
27 & 7.4 & 0.0199 & 0.054 & 10 & 74.9 & -0.8 & 0.38\\
28 & 7.5 & 0.0203 & 0.056 & 10 & 28.6 & 0.1 & 0.14\\ 
29 & 7.5 & 0.0203 & 0.056 & 10 & 90.4 & -3.0 & 0.44\\ 
30 & 8.0 & 0.0221 & 0.067 & 11 & 93.9 & 3.2 & 0.38\\
31 & 8.1 & 0.0225 & 0.069 & 11 & 78.3 & 1.3 & 0.31\\ [1ex] 
\hline 
\end{tabu} 
\end{table} 

In order to estimate the bed-load transport rate, we averaged the instantaneous velocity fields and considered the total number of moving grains in each instantaneous field to be $20$ times the number of moving black spheres $N_p$, due to the bed composition. The bed-load transport rate was then computed as

\begin{equation}
Q_B\,=\,d\, v_x\, L\, \frac{20N_p}{A_{field}} \frac{\pi d^2}{4}
\label{exp_transp_rate}
\end{equation}

\noindent where $Q_B$ is the volumetric bed-load transport rate, $L$ is the channel width, and $A_{field}$ is the area filmed by the camera. Table \ref{table:table2} presents the water flow rate $Q$, Reynolds number $Re$, particle Reynolds number $Re_*$, Shields number $\theta$, bed-load transport rate $Q_B$, dimensionless bed-load transport rate $\phi_B$, estimated threshold Shields number $\theta_{th}$, and a reference threshold Shields number $\theta_{th,ref}$ for different test runs.

\begin{table}[ht]
\caption{Test run, water flow rate $Q$, Reynolds number $Re$, Shields number $\theta$, particle Reynolds number $Re_*$, bed-load transport rate $Q_B$, dimensionless bed-load transport rate $\phi_B$, estimated threshold Shields number $\theta_{th}$, and a reference threshold Shields number $\theta_{th,ref}$.}
\label{table:table2}
\centering 
\begin{tabu}{c c c c c c c c c}  
\hline\hline 
run & $Q$ & $Re$ & $\theta$ & $Re_*$ & $Q_B$ & $\phi_B$ & $\theta_{th}$ & $\theta_{th,ref}$\\
$\cdots$ & $m^3/h$ & $\cdots$ & $\cdots$ & $\cdots$ & $m^3/s$ & $\cdots$ & $\cdots$ & $\cdots$\\ [0.5ex] 
\hline 
1 & 4.8 & $1.7\cdot 10^{4}$ & 0.014 & 5 & $1.89 \times 10^{-9}$ & $2.7\cdot 10^{-4}$ & 0.010 & 0.023\\ 
2 & 5.7 & $2.0\cdot 10^{4}$ & 0.025 & 7 & $5.34 \times 10^{-9}$ & $0.8\cdot 10^{-3}$ & 0.020 & 0.032\\ 
3 & 6.2 & $2.2\cdot 10^{4}$ & 0.032 & 8 & $12.0 \times 10^{-9}$ & $1.7\cdot 10^{-3}$ & 0.025 & 0.033\\ 
4 & 6.2 & $2.2\cdot 10^{4}$ & 0.032 & 8 & $4.21 \times 10^{-9}$ & $0.6\cdot 10^{-3}$ & 0.025 & 0.051\\ 
5 & 6.3 & $2.2\cdot 10^{4}$ & 0.033 & 8 & $7.24 \times 10^{-9}$ & $1.1\cdot 10^{-3}$ & 0.025 & 0.037\\ 
6 & 6.3 & $2.2\cdot 10^{4}$ & 0.034 & 8 & $21.3 \times 10^{-9}$ & $3.1\cdot 10^{-3}$ & 0.025 & 0.048\\
7 & 6.4 & $2.2\cdot 10^{4}$ & 0.035 & 8 & $6.4 \times 10^{-9}$ & $0.9\cdot 10^{-3}$ & 0.025 & 0.051\\
8 & 6.4 & $2.2\cdot 10^{4}$ & 0.035 & 8 & $15.3 \times 10^{-9}$ & $2.2\cdot 10^{-3}$ & 0.025 & 0.042\\  
9 & 6.5 & $2.3\cdot 10^{4}$ & 0.037 & 8 & $7.5 \times 10^{-9}$ & $1.1\cdot 10^{-3}$ & 0.025 & 0.047\\ 
10 & 6.5 & $2.3\cdot 10^{4}$ & 0.037 & 8 & $11.8 \times 10^{-9}$ & $1.7\cdot 10^{-3}$ & 0.025 & 0.053\\ 
11 & 6.5 & $2.3\cdot 10^{4}$ & 0.037 & 8 & $14.3 \times 10^{-9}$ & $2.1\cdot 10^{-3}$ & 0.025 & 0.055\\
12 & 6.6 & $2.3\cdot 10^{4}$ & 0.039 & 8 & $20.9 \times 10^{-9}$ & $3.0\cdot 10^{-3}$ & 0.025 & 0.027\\ 
13 & 6.6 & $2.3\cdot 10^{4}$ & 0.039 & 8 & $9.1 \times 10^{-9}$ & $1.3\cdot 10^{-3}$ & 0.025 & 0.044\\ 
14 & 6.6 & $2.3\cdot 10^{4}$ & 0.039 & 8 & $8.4 \times 10^{-9}$ & $1.2\cdot 10^{-3}$ & 0.025 & 0.044\\ 
15 & 6.7 & $2.3\cdot 10^{4}$ & 0.041 & 9 & $16.8 \times 10^{-9}$ & $2.4\cdot 10^{-3}$ & 0.030 & 0.033\\ 
16 & 6.7 & $2.3\cdot 10^{4}$ & 0.041 & 9 & $8.7 \times 10^{-9}$ & $1.3\cdot 10^{-3}$ & 0.030 & 0.044\\ 
17 & 6.8 & $2.4\cdot 10^{4}$ & 0.042 & 9 & $23.7 \times 10^{-9}$ & $3.5\cdot 10^{-3}$ & 0.030 & 0.060\\
18 & 6.9 & $2.4\cdot 10^{4}$ & 0.044 & 9 & $8.0 \times 10^{-9}$ & $1.2\cdot 10^{-3}$ & 0.035 & 0.064\\
19 & 7.0 & $2.4\cdot 10^{4}$ & 0.046 & 9 & $9.31 \times 10^{-9}$ & $1.4\cdot 10^{-3}$ & 0.035 & 0.068\\ 
20 & 7.0 & $2.4\cdot 10^{4}$ & 0.046 & 9 & $15.6 \times 10^{-9}$ & $2.3\cdot 10^{-3}$ & 0.035 & 0.066\\ 
21 & 7.0 & $2.4\cdot 10^{4}$ & 0.046 & 9 & $11.3 \times 10^{-9}$ & $1.6\cdot 10^{-3}$ & 0.035 & 0.054\\ 
22 & 7.1 & $2.5\cdot 10^{4}$ & 0.048 & 9 & $48.2 \times 10^{-9}$ & $7.0\cdot 10^{-3}$ & 0.035 & 0.056\\ 
23 & 7.1 & $2.5\cdot 10^{4}$ & 0.048 & 9 & $18.0 \times 10^{-9}$ & $2.6\cdot 10^{-3}$ & 0.035 & 0.041\\ 
24 & 7.2 & $2.5\cdot 10^{4}$ & 0.050 & 10 & $13.8 \times 10^{-9}$ & $2.0\cdot 10^{-3}$ & 0.040 & 0.059\\ 
25 & 7.3 & $2.5\cdot 10^{4}$ & 0.051 & 10 & $23.3 \times 10^{-9}$ & $3.4\cdot 10^{-3}$ & 0.040 & 0.042\\ 
26 & 7.3 & $2.5\cdot 10^{4}$ & 0.051 & 10 & $24.8 \times 10^{-9}$ & $3.6\cdot 10^{-3}$ & 0.040 & 0.059\\ 
27 & 7.4 & $2.6\cdot 10^{4}$ & 0.054 & 10 & $17.7 \times 10^{-9}$ & $2.6\cdot 10^{-3}$ & 0.045 & 0.050\\
28 & 7.5 & $2.6\cdot 10^{4}$ & 0.056 & 10 & $15.7 \times 10^{-9}$ & $2.3\cdot 10^{-3}$ & 0.045 & 0.090\\ 
29 & 7.5 & $2.6\cdot 10^{4}$ & 0.056 & 10 & $22.2 \times 10^{-9}$ & $3.2\cdot 10^{-3}$ & 0.045 & 0.043\\ 
30 & 8.0 & $2.8\cdot 10^{4}$ & 0.067 & 11 & $34.6 \times 10^{-9}$ & $5.0\cdot 10^{-3}$ & 0.050 & 0.054\\ 
31 & 8.1 & $2.8\cdot 10^{4}$ & 0.069 & 11 & $40.0 \times 10^{-9}$ & $5.8\cdot 10^{-3}$ & 0.050 & 0.068\\ [1ex] 
\hline 
\end{tabu} 
\end{table} 

Next, we compare the transport rates estimated from the instantaneous fields with some semi-empirical expressions. Given the large uncertainty concerning the threshold Shields number, explained in the following, the objective of the comparisons with the semi-empirical expressions is to observe the tendencies of the measured velocities and surface densities. One of the most frequently employed expressions is that of Meyer-Peter and  M\"{u}ller (1948) \cite{Mueller}, given by $\phi_B=a(\theta-\theta_{th})^{3/2}$, where $a=4$ in the present case, because ripples were absent in our tests. Other expressions employed here are those proposed by Bagnold (1956) \cite{Bagnold_3}, by Lettau and Lettau (1978) \cite{Lettau}, and by Abrahams and Gao (2006) \cite{Abrahams_gao}. In these expressions, $A$ is a constant that depends on the Reynolds number and is taken here as $5.5$ \cite{Bagnold_3}, while $C_L$ is a constant to be adjusted and taken here as $12$ \cite{Lettau}. The major difficulty in employing these semi-empirical expressions concerns the threshold Shields number, $\theta_{th}$, because values in the literature present large dispersions.

As shown by Charru et al. (2004) \cite{Charru_1}, the threshold shear stress may grow over time because of an increase in bed compactness. They proposed that the increase in bed compactness is caused by the grains' movement, so that it occurs faster under higher shear stresses. Given the sequence in which the present tests were performed, the values of $\theta_{th}$ are expected to increase with the water flow rate. Fernandez Luque and van Beek (1976) \cite{Fernandez_luque} proposed that the threshold shear stress operating at the topmost layer of the granular bed increases with the applied shear stress. The reason is that the feedback effect increases with the fluid shear on the bed surface. This proposition also indicates that in our tests the values of $\theta_{th}$ increase with the water flow rate.

In the present experiments, it was clear that $\theta_{th}$ was smaller for the smaller flow rates, with the measured Shields number being below the usually employed $\theta_{th} = 0.04$ (Tab. \ref{table:table2}). In order to have an excess in shear stress, and then bed load, $\theta_{th}$ must be smaller than $0.04$ for the lower water flow rates. For this reason, the values of $\theta_{th}$ were varied between $0.010$ and $0.050$, the values expected for loose beds\cite{Raudkivi_1, Yalin_2, Buffington_1}. The corresponding values, which were increased with the fluid shear by steps of $0.005$, are presented in Tab. \ref{table:table2}. Only as a means of comparison, the formula proposed by Fernandez Luque and van Beek (1976) \cite{Fernandez_luque} for the mean velocity of grains, $v_x = 11.5 (u_* - 0.7u_{*,th})$, was used to estimate, from the measured $v_x$ values, the threshold shear velocity $u_{*,th}$, and then threshold Shields number. The values of threshold Shields number obtained from the Fernandez Luque and van Beek's (1976) \cite{Fernandez_luque} formula, $\theta_{th,ref}$, are also presented in Tab. \ref{table:table2}. It can be seen that the values are different for each run, and that there is a tendency to increase with the water flow rate.

\begin{figure}[!ht]
   \begin{minipage}[c]{.5\linewidth}
    \begin{center}
     \includegraphics[width=0.99\columnwidth]{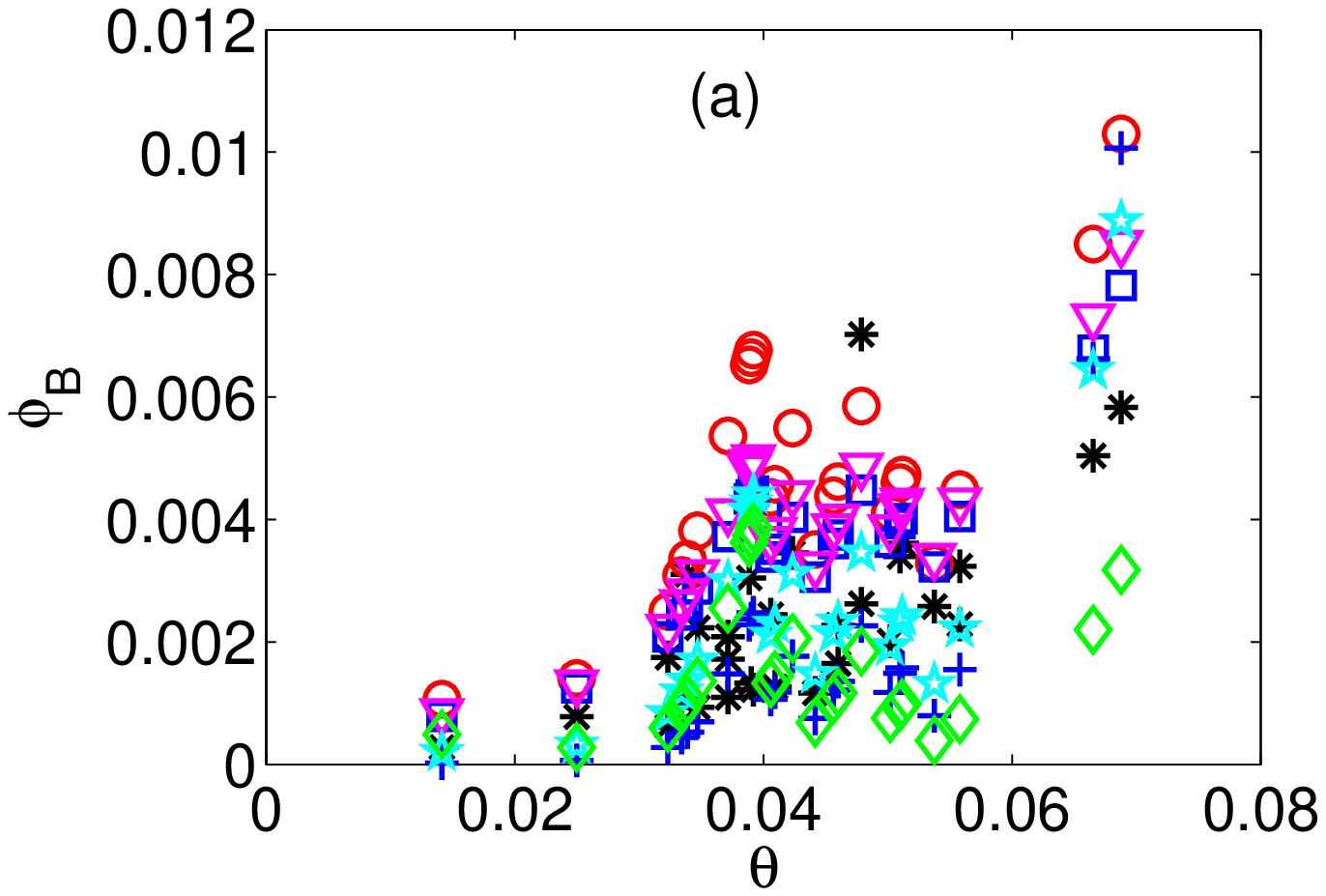}
    \end{center}
   \end{minipage} \hfill
   \begin{minipage}[c]{.5\linewidth}
    \begin{center}
      \includegraphics[width=0.99\columnwidth]{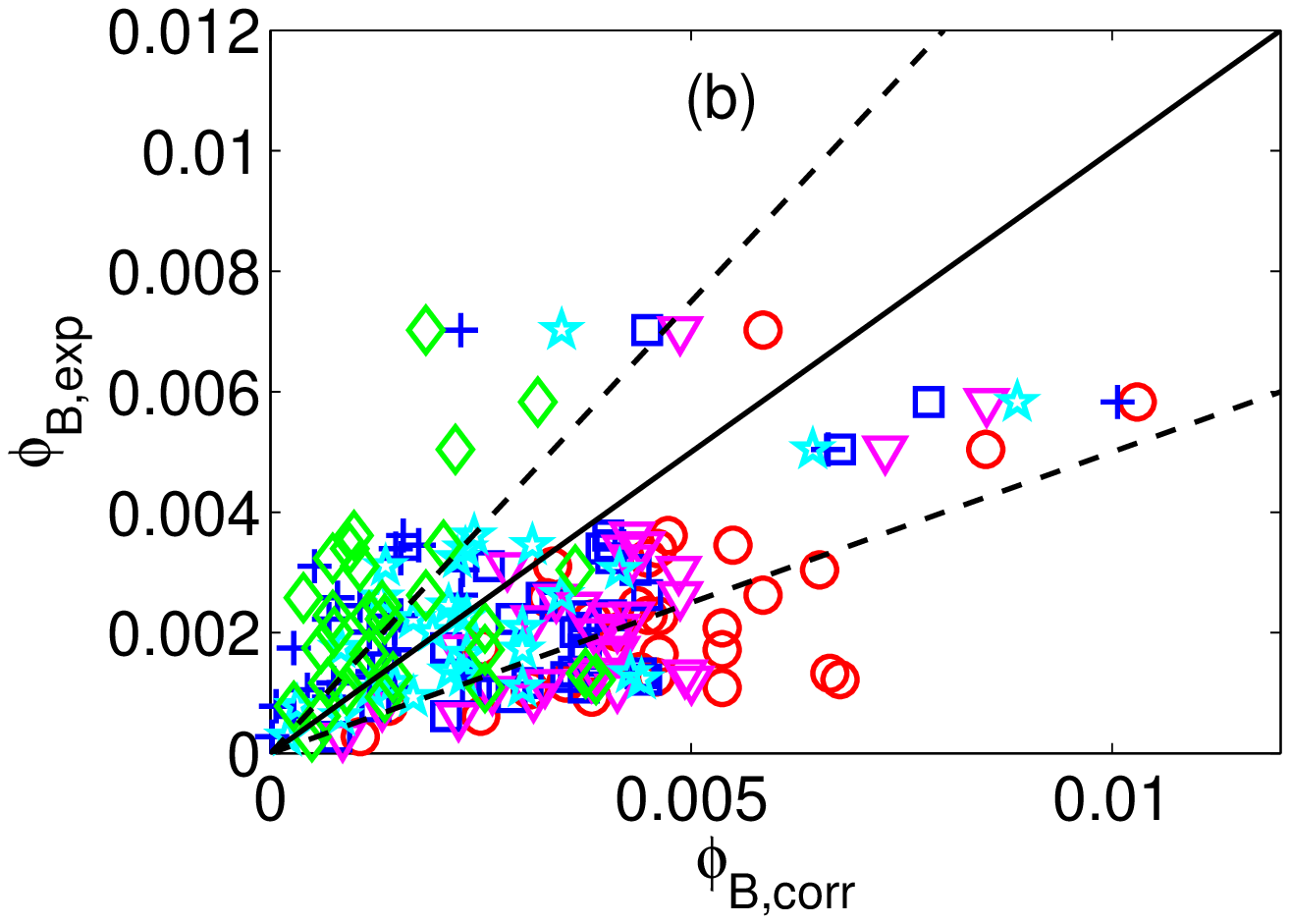}
    \end{center}
   \end{minipage}
\caption{(a) Dimensionless bed-load transport rate $\phi_B$ as a function of Shields number $\theta$. (b) Measured bed-load transport rate (dimensionless) $\phi_{B,exp}$ as a function of semi-empirical expressions for the bed-load transport rate (dimensionless) $\phi_{B,corr}$. The asterisks, circles, squares, inverted triangles, lozenges, crosses, and stars correspond to measured values, and to the Meyer-Peter and  M\"{u}ller (1948) \cite{Mueller}, Bagnold (1956) \cite{Bagnold_3}, Lettau and Lettau (1978) \cite{Lettau},  Abrahams and Gao (2006) \cite{Abrahams_gao}, Franklin (2008) \cite{Franklin_3}, and Franklin and Charru (2011) \cite{Franklin_9} expressions, respectively. The continuous line corresponds to $\phi_{B,exp} = \phi_{B,corr}$, and the two dashed lines correspond to $\phi_{B,exp} = 0.5 \phi_{B,corr}$ and to $\phi_{B,exp} = 1.5 \phi_{B,corr}$}
\label{fig:QB}
\end{figure}

Figure \ref{fig:QB}(a) presents the dimensionless bed-load transport rate $\phi_B$ as a function of the Shields number $\theta$. The asterisks, circles, squares, inverted triangles, lozenges, crosses, and stars correspond to measured values, and to the Meyer-Peter and  M\"{u}ller (1948) \cite{Mueller}, Bagnold (1956) \cite{Bagnold_3}, Lettau and Lettau (1978) \cite{Lettau},  Abrahams and Gao (2006) \cite{Abrahams_gao}, Franklin (2008) \cite{Franklin_3}, and Franklin and Charru (2011) \cite{Franklin_9} expressions, respectively. Figure \ref{fig:QB}(b) presents the measured bed-load transport rate (dimensionless) $\phi_{B,exp}$ as a function of semi-empirical expressions for the bed-load transport rate (dimensionless) $\phi_{B,corr}$. The circles, squares, inverted triangles, lozenges, crosses, and stars correspond to the Meyer-Peter and  M\"{u}ller (1948) \cite{Mueller}, Bagnold (1956) \cite{Bagnold_3}, Lettau and Lettau (1978) \cite{Lettau},  Abrahams and Gao (2006) \cite{Abrahams_gao}, Franklin (2008) \cite{Franklin_3}, and Franklin and Charru (2011) \cite{Franklin_9} expressions, respectively. The continuous line corresponds to $\phi_{B,exp} = \phi_{B,corr}$, and the two dashed lines correspond to $\phi_{B,exp} = 0.5 \phi_{B,corr}$ and to $\phi_{B,exp} = 1.5 \phi_{B,corr}$. In the case of the Franklin and Charru (2011) \cite{Franklin_9} expression, the multiplicative pre-factor was altered from $34$ to $5$ in order to fit this specific expression with our experimental data. This expression was obtained for the transport rate over a dune, and it super-estimates the transport rate over flat beds.

There are several expressions for the bed-load transport rate, many of them with the same functional form: power functions of the Shields number and the Threshold Shields number. Usually, the differences between these expressions are in the multiplicative pre-factor and the power coefficient. Because the transport rate equations are empirical and semi-empirical laws, experimental uncertainties are included in both the multiplicative pre-factor and the power coefficient. In addition, high uncertainties are present in the determination of the threshold shear stress because it depends on the surface density of the moving grains, which may change over time, as shown by Charru et al. (2004) \cite{Charru_1}. For all these reasons, the different existing models give different trends, and there is a lack of universality. The transport rate results obtained from the velocity fields in the present study lie among the results predicted by different models. Given the uncertainties and differences related to the bed-load transport rate equations, the experimentally determined bed-load transport rates are in good agreement with semi-empirical expressions proposed in the literature, corroborating the measurements of individual displacements and the determination of velocity fields.

\subsection{Lagrangian framework}

\begin{figure}[!ht]
   \begin{minipage}[c]{.49\linewidth}
    \begin{center}
     \includegraphics[width=0.99\columnwidth]{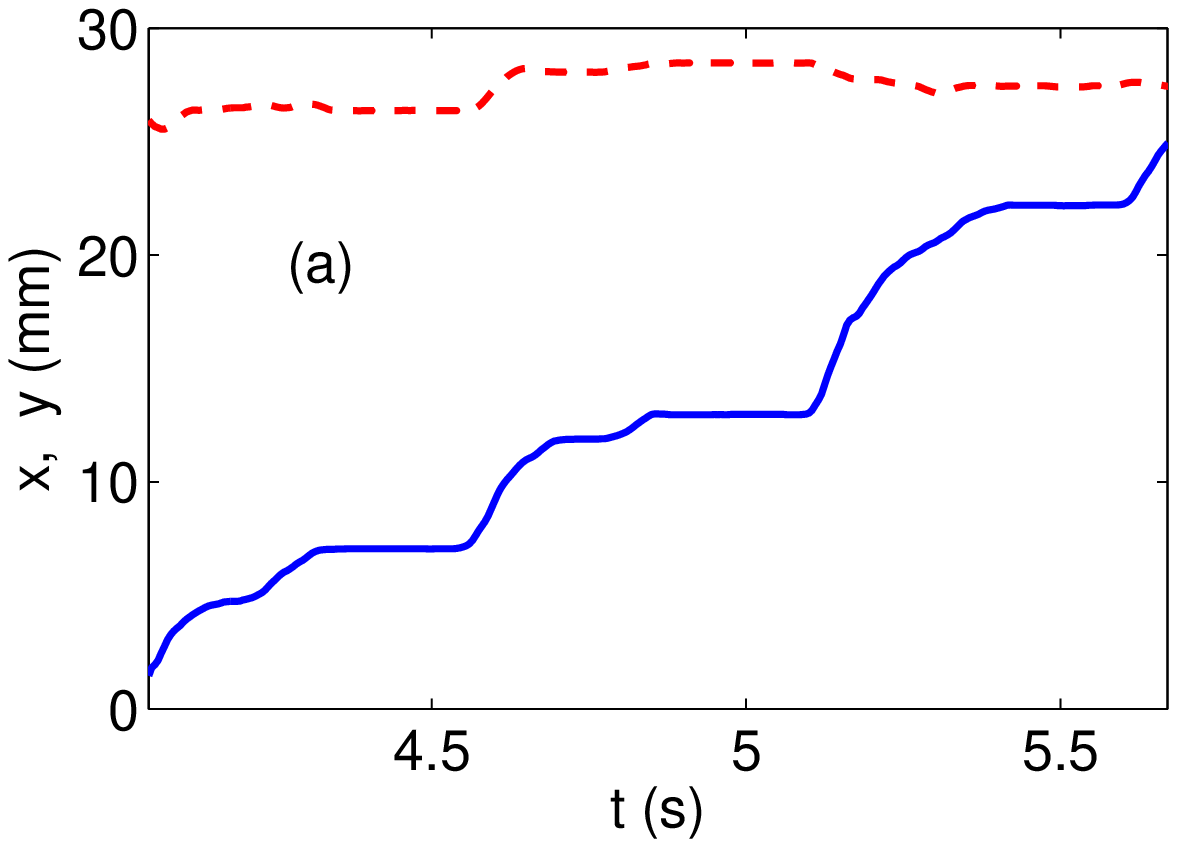}
    \end{center}
   \end{minipage} \hfill
   \begin{minipage}[c]{.49\linewidth}
    \begin{center}
      \includegraphics[width=0.99\columnwidth]{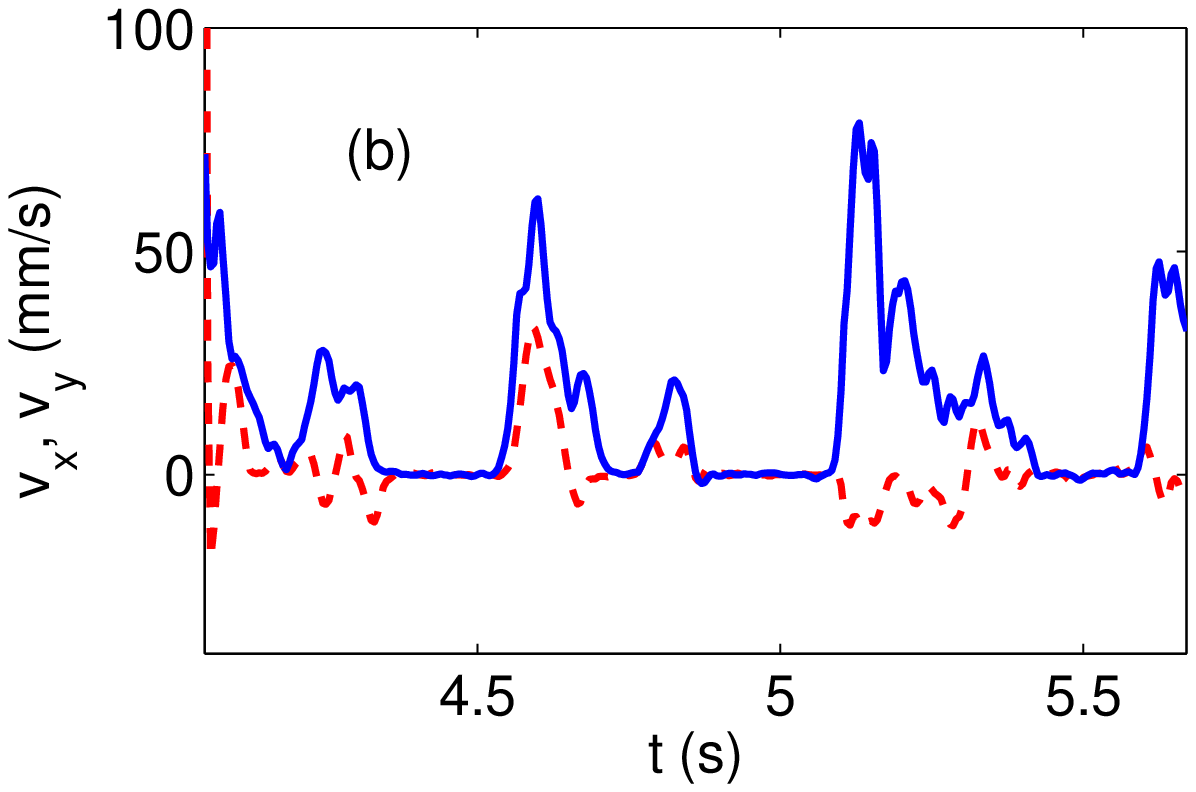}
    \end{center}
   \end{minipage}
\caption{(a) Displacement of an individual grain. (b) Velocity of an individual grain. The continuous and the dashed curves correspond to the $x$ and $y$ components, respectively. $Q = 6.8 m^3/h$ (run 17).}
\label{fig:lagrangian}
\end{figure}

The trajectories of individual grains were determined by tracking some of the black spheres along the images of a given run. Figure \ref{fig:lagrangian}(a) presents the $x$ and $y$ position components of a given glass sphere as function of time $t$, and Fig. \ref{fig:lagrangian}(b) presents the $x$ and $y$ velocity components of a given glass sphere as function of time $t$, both for $Q = 6.8 m^3/h$ (run 17). In these figures, $x$ and $y$ are, respectively, the longitudinal and the transverse components of the position vector, and $v_x$ and $v_y$ are, respectively, the longitudinal and the transverse components of the velocity vector. The continuous and the dashed curves correspond to the $x$ and $y$ components, respectively.

Figure \ref{fig:lagrangian}(a) shows that the displacement of individual grains is intermittent, with periods of acceleration, deceleration and at rest, which is corroborated by Fig. \ref{fig:lagrangian}(b). For this water flow rate, the typical grain displacement is $\approx 10d$.

\begin{table}[!ht]
  \begin{center}
    \caption{Test run, water flow rate $Q$, mean longitudinal displacement of grains $\Delta x$, standard deviation of longitudinal displacement $\sigma_{\Delta x}$, mean displacement velocity $U_{d}$, standard deviation of displacement velocity $\sigma_{U_d}$, normalized longitudinal displacement $\Delta x^{ad}$, normalized displacement time $t_d^{ad}$, normalized mean displacement velocity $ U_{d}^{ad}$.}
		\label{tab_desl}
    \begin{tabu}{c c c c c c c c c}		
      \hline\hline 
			run & $Q$ & $\Delta x$ & $\sigma_{\Delta x}$ & $U_{d}$ & $\sigma_{U_d}$ & $\Delta x^{ad}$ & $t^{ad}$ & $ U_{d}^{ad}$\\
			$\cdots$ & $m^3/h$ & $mm$ & $mm$ & $mm/s$ & $mm/s$ & $\cdots$ & $\cdots$ & $\cdots$\\
      \hline
      1 & $4.8$ & $4.1$ & $2.0$ & $21.1$ & $2.7$ & $8.2$ & $28.2$ & $0.33$\\
      
			2 & $5.7$ & $5.6$ & $3.8$ & $44.9$ & $14.3$ & $11.2$ & $18.3$ & $0.48$\\
			
			6 & $6.3$ & $9.4$ & $3.7$ & $39.0$ & $8.0$ & $16.8$ & $35.0$ & $0.35$\\
      
			10 & $6.5$ & $9.9$ & $4.3$ & $40.9$ & $7.4$ & $19.8$ & $35.4$ & $0.33$\\
			
			11 & $6.5$ & $5.8$ & $2.3$ & $36.0$ & $8.0$ & $11.6$ & $23.7$ & $0.29$\\
			
			17 & $6.8$ & $6.9$ & $1.7$ & $45.0$ & $13.9$ & $13.8$ & $22.5$ & $0.30$\\
      
			19 & $7.0$ & $7.4$ & $4.8$ & $66.0$ & $15.1$ & $14.8$ & $16.5$ & $0.39$\\
			
			22 & $7.1$ & $11.3$ & $5.7$ & $45.3$ & $11.2$ & $22.6$ & $36.9$ & $0.25$\\
      
			29 & $7.5$ & $8.8$ & $5.3$ & $58.0$ & $19.1$ & $17.6$ & $22.3$ & $0.28$\\
       
			31 & $8.1$ & $6.5$ & $3.6$ & $59.5$ & $16.4$ & $13.0$ & $15.9$ & $0.25$\\
      \hline 
    \end{tabu}
  \end{center}
\end{table}

The longitudinal displacement of some moving black spheres was computed as the distance traveled by the sphere between periods at rest, and the mean longitudinal displacement $\Delta x$ was determined by averaging the longitudinal displacement of black spheres for individual test runs. Therefore, $\Delta x$ corresponds to the typical longitudinal distance traveled by grains between periods at rest for a given water flow condition. Table \ref{tab_desl} shows $\Delta x$ as well as the standard deviation of the longitudinal displacement of grains $\sigma_{\Delta x}$. The mean longitudinal displacement normalized by the grain diameter $\Delta x^{ad}=\Delta x/d$ is also shown in Tab. \ref{tab_desl}, and it is seen to vary between $8.2$ and $22.6$.

The longitudinal velocity of moving grains was computed by considering only the periods for which each black sphere was moving, i.e., the periods of acceleration and deceleration. The mean displacement velocity $U_d$ was then obtained by averaging all the computed longitudinal velocities, for individual test runs. It is presented in Tab. \ref{tab_desl} together with the standard deviation of the longitudinal velocity, $\sigma_{U_d}$. The mean displacement velocity was normalized by the shear rate and the grain diameter $U_d^{ad} = \nu U_d(u_*^2 d)^{-1}$ and is shown in Tab. \ref{tab_desl}. In all tested conditions,  $U_d^{ad} = ord(0.1)$, in agreement with the Eulerian framework and with Charru et al. (2004) \cite{Charru_1}.

The mean displacement time was computed as $t_d = \Delta x/U_d$, and it corresponds to the time taken by individual grains to travel the typical displacement distance $\Delta x$ at the typical displacement velocity $U_d$. The mean displacement time was normalized by the grain diameter and the settling velocity $t_d^{ad} = t_d U_s/d$, where $U_s$ is the settling velocity of a single grain. The normalized displacement time is shown in Tab. \ref{tab_desl}. The settling velocity was computed by the Schiller--Naumann correlation \cite{Clift}. The computed values of $t_d^{ad}$ are between $15.9$ and $36.9$, in agreement with Charru et al. (2004) \cite{Charru_1}.

As a final comment, we note the importance of the trajectories of grains and velocity fields of a bed-load layer for determining the morphology of granular beds sheared by fluids. In order to determine the stability of a granular bed and to understand the evolution of bed morphology, the local variation of bed-load transport rate is crucial \cite{Charru_3, Franklin_4}; therefore, it is important to know the grains trajectories and velocity fields. While in the aeolian turbulent case the trajectories of salting grains is known \cite{Sauermann_4, Sauermann_3} and the growth rate, lengths and celerities of sand dunes understood \cite{Kroy_C, Andreotti_1, Andreotti_2, Parteli, Melo, Luna}, in the case of turbulent water flows there is a lack of data.

\section{Conclusions}
\label{section:conclusions}

The study presented in this paper was devoted to the displacements of grains within a moving granular layer under a turbulent water flow. A granular bed of known granulometry was sheared by fully developed turbulent water flows. The present experiments were performed close to incipient bed load, a case for which experimental data on grains velocities are scarce, and the granular bed remained flat in the course of all tests. Distinct from previous experiments, an entrance length of $40$ hydraulic diameters assured that the water stream over the loose bed was fully developed. At different flow rates, the displacements of grains were filmed using a high-speed camera, from which the trajectories and velocities of individual grains were computed.

A numerical code for image treatment was developed, and the images were post-processed in an Eulerian framework in order to obtain the grains' displacements and velocity fields. The results showed that the normalized longitudinal displacement velocity, $v_x^{ad} = \nu v_x/(u_*^2d)$, is of the order of $0.1$, in agreement with Charru et al.'s (2004) results \cite{Charru_1}, although their experiments addressed viscous flows. The software Motion Studio was used to track individual grains in a Lagragian framework in some of the experiments. The experimental results showed that individual grains have an intermittent motion, with periods of acceleration, deceleration and at rest. The typical traveled distances, displacement times and velocities were computed based on the trajectories of individual grains. The mean longitudinal displacement normalized by the grain diameter, $\Delta x^{ad}=\Delta x/d$, the  mean displacement velocity normalized by the shear rate and the grain diameter, $U_d^{ad} = \nu U_d(u_*^2 d)^{-1}$, and the mean displacement time normalized by the grain diameter and the settling velocity, $t_d^{ad} = t_d U_s/d$, were found to be of the order of $10$, $0.1$, and $10$, respectively, in agreement with the Eulerian framework and with Charru et al. (2004) \cite{Charru_1}.

The bed-load transport rate was estimated on the basis of the velocity fields and correlated with water flow conditions. The experimentally determined bed-load transport rates were compared with semi-empirical expressions proposed in the literature. Because we did not measure the threshold Shields number, which seems to vary along the different test runs as all tests were close to incipient conditions, and given the large uncertainties concerning the threshold Shields number, the objective of the comparisons with the semi-empirical expressions was only to verify the tendencies of the measured velocities and surface densities. Based on armouring arguments, we varied the threshold Shields number between $0.010$ and $0.050$. Even with this simple estimation, the results obtained from semi-empirical expressions were found to have a reasonable agreement with the ones obtained from measured velocities and surface densities.

\section{Acknowledgments}

\begin{sloppypar}
The authors are grateful to FAPESP (grant no. 2012/19562-6), to CNPq (grant no. 471391/2013-1), to FAEPEX/UNICAMP (conv. 519.292, projects AP0008/2013 and 0201/14), and to CAPES for the provided financial support. The authors thank Rodolfo M. Perissinotto for the help with the experimental device. 
\end{sloppypar}






\bibliography{references}

\begin{thebibliography}{10}
\expandafter\ifx\csname url\endcsname\relax
  \def\url#1{\texttt{#1}}\fi
\expandafter\ifx\csname urlprefix\endcsname\relax\def\urlprefix{URL }\fi

\bibitem{Franklin_4}
E.~M. Franklin, Initial instabilities of a granular bed sheared by a turbulent
  liquid flow: length-scale determination, J. Braz. Soc. Mech. Sci. Eng. 32~(4)
  (2010) 460--467.

\bibitem{Franklin_6}
E.~M. Franklin, Linear and nonlinear instabilities of a granular bed:
  determination of the scales of ripples and dunes in rivers, Appl. Math.
  Model. 36 (2012) 1057--1067.

\bibitem{Franklin_8}
E.~M. Franklin, F.~T. Figueiredo, E.~S. Rosa, The feedback effect caused by bed
  load on a turbulent liquid flow, J. Braz. Soc. Mech. Sci. Eng. 36 (2014)
  725--736.

\bibitem{Raudkivi_1}
A.~J. Raudkivi, Loose boundary hydraulics, 1st Edition, Pergamon Press, 1976.

\bibitem{Fernandez_luque}
R.~Fernandez~Luque, R.~van Beek, Erosion and transport of bed-load sediment, J.
  Hydraul. Res. 14 (2010) 127--144.

\bibitem{Charru_1}
F.~Charru, H.~Mouilleron-Arnould, O.~Eiff, Erosion and deposition of particles
  on a bed sheared by a viscous flow, J. Fluid Mech. 519 (2004) 55--80.

\bibitem{Charru_4}
F.~Charru, H.~Mouilleron-Arnould, O.~Eiff, Inside the moving layer of a sheared
  granular bed, J. Fluid Mech. 629 (2009) 229--239.

\bibitem{Lajeunesse_1}
E.~Lajeunesse, L.~Malverti, F.~Charru, Bed load transport in turbulent flow at
  the grain scale: {E}xperiments and modeling, J. Geophys. Res. 115~(F4),
  f04001.

\bibitem{Mueller}
E.~Meyer-Peter, R.~M\"{u}ller, Formulas for bed-load transport, in: Proc. 2nd
  Meeting of International Association for Hydraulic Research, 1948, pp.
  39--64.

\bibitem{Wong_parker}
M.~Wong, G.~Parker, Reanalysis and correction of bed-load relation of
  {M}eyer-{P}eter and {M}\"{u}ller using their own database, J. Hydraul. Eng.
  132 (2006) 1159--1168.

\bibitem{Bagnold_3}
R.~A. Bagnold, The flow of cohesionless grains in fluids, Philos. Trans. R.
  Soc. Lond. Ser. A 249 (1956) 235--297.

\bibitem{Bagnold_4}
R.~A. Bagnold, The nature of saltation and of 'bed-load' transport in water,
  Philos. Trans. R. Soc. Lond. Ser. A 334 (1973) 473--504.

\bibitem{Lettau}
K.~Lettau, H.~Lettau, Experimental and micrometeorological field studies of
  dune migration, in: H.~H. Lettau, K.~Lettau (Eds.), Exploring the world's
  driest climate, University of Wisconsin, Madison, Center for Climatic
  Research, Univ. Wisconsin, 1978, pp. 110--147.

\bibitem{Abrahams_gao}
A.~D. Abrahams, P.~Gao, A bed-load transport model for rough turbulent
  open-channel flows on plane beds, Earth Surf. Processes and Landforms 31
  (2006) 910--928.

\bibitem{Franklin_3}
E.~M. Franklin, Dynamique de dunes isol\'ees dans un \'ecoulement cisaill\'e,
  Ph.D. thesis, Universit\'e de Toulouse (2008).

\bibitem{Franklin_9}
E.~M. Franklin, F.~Charru, Subaqueous barchan dunes in turbulent shear flow.
  {P}art 1. {D}une motion, J. Fluid Mech. 675 (2011) 199--222.

\bibitem{Gao}
P.~Gao, Validation and implications of an energy-based bedload transport
  equation, Sedimentology 59~(6) (2012) 1926--1935.

\bibitem{Sonka}
M.~Sonka, V.~Hlavac, R.~Boyle, Image processing, analysis, and machine vision,
  Chapman and Hall, 1993.

\bibitem{Yalin_2}
M.~S. Yalin, E.~Karahan, Inception of sediment transport, J. Hydraul. Div. HY11
  (1979) 1433--1443.

\bibitem{Buffington_1}
J.~M. Buffington, D.~R. Montgomery, A systematic analysis of eight decades of
  incipient motion studies, with special reference to gravel-bedded rivers,
  Water Resour. Res. 33 (1997) 1993--2029.

\bibitem{Clift}
R.~Clift, J.~Grace, M.~Weber, Bubbles, drops and particles, 1st Edition, Dover
  Publications, 2005.

\bibitem{Charru_3}
F.~Charru, Selection of the ripple length on a granular bed sheared by a liquid
  flow, Phys. of Fluids 18~(121508).

\bibitem{Sauermann_4}
G.~Sauermann, K.~Kroy, H.~J. Herrmann, Continuum saltation model for sand
  dunes, Phys. Rev. E 64~(031305).

\bibitem{Sauermann_3}
G.~Sauermann, J.~S. Andrade, L.~P. Maia, U.~M.~S. Costa, A.~D. Ara\'ujo, H.~J.
  Herrmann, Wind velocity and sand transport on a barchan dune, Geomorphology
  54 (2003) 245--255.

\bibitem{Kroy_C}
K.~Kroy, G.~Sauermann, H.~J. Herrmann, Minimal model for sand dunes, Phys. Rev.
  Lett. 88~(054301).

\bibitem{Andreotti_1}
B.~Andreotti, P.~Claudin, S.~Douady, Selection of dune shapes and velocities.
  part 1: {D}ynamics of sand, wind and barchans, Eur. Phys. J. B 28 (2002)
  321--329.

\bibitem{Andreotti_2}
B.~Andreotti, P.~Claudin, S.~Douady, Selection of dune shapes and velocities.
  part 2: {A} two-dimensional model, Eur. Phys. J. B 28 (2002) 341--352.

\bibitem{Parteli}
E.~J.~R. Parteli, O.~Dur\'an, H.~J. Herrmann, Minimal size of a barchan dune,
  Phys. Rev. E 75~(011301).

\bibitem{Melo}
H.~P.~M. Melo, E.~J.~R. Parteli, J.~S. Andrade, H.~J. Herrmann, Linear
  stability analysis of transverse dunes, Physica A 391~(20) (2012) 4606 --
  4614.

\bibitem{Luna}
M.~C. M.~M. Luna, E.~J.~R. Parteli, H.~J. Herrmann, Model for a dune field with
  an exposed water table, Geomorphology 159–160 (2012) 169 -- 177.

\end{thebibliography}
\bibliographystyle{elsart-num}







\end{document}